\pdfoutput=1
 
\documentclass[nofootinbib,twocolumn,showpacs,aps,prd,superscriptaddress,amsmath,amssymb,floatfix,twoside]{revtex4}
\usepackage{graphicx}
\usepackage{bm}
\usepackage{color}
\usepackage{multirow}
\usepackage{bookmark}

\begin{document}

\title{Computer simulation of detecting system
of compact positron-emission tomograph
based on scintillator-photodiode detectors}

\author{A.S.~Fomin}
\email[]{fomax.ua@gmail.com}
\affiliation{NSC Kharkiv Institute of Physics and Technology, 61108 Kharkiv, Ukraine}
\affiliation{V.N.~Karazin Kharkiv National University, 61022 Kharkiv, Ukraine}

\date{2004, 5 July}

\begin{abstract}

We present the original computer code for the simulation of multi-element detection system of the compact positron-emission tomograph based on a scintillator-photodiode type of detection elements. The use of such type of detection elements allows obtaining a high spatial resolution at a relatively small total size of the tomograph. This program gives an opportunity to choose the optimal geometry of detection system depending on the parameters of its elements, and analyze the efficiency of different image reconstruction algorithms. The handy interface for the designed compact positron-emission tomograph operation has been created.

\end{abstract}


\maketitle

\setcounter{footnote}{0}

%
%%%
%%%%%%%%%%%%%%%%%%%%%%%%%%%%%%%%%%%%%%%%%%%

\section{\label{sec:introduction} Introduction}

It is well known that the positron-emission tomography (PET) is one of the most powerful and quickly developing diagnostic techniques, which allows carrying out the measurements of distributions of atoms traced by the $\beta^+$-active nuclide in a living organism~\cite{Lecoq}. The possibility of wide adoption of this technique in medical diagnostics appears due to the existence of four neutron-lack isotopes $^{11}$C, $^{15}$O, $^{13}$N and $^{17}$F of the basic elements that belong to all biological tissues and take part in nearly all biological processes in the organism.

In comparison with ordinary tomography PET has a medium (about 5--6~mm) spatial resolution and gives an opportunity of decreasing the radiation dose on the patient organism due to a high efficiency and by using ultra short life-time isotopes (USLI) with half-value period $\tau \approx$~2--110~minutes (see Table~\ref{tab:isotopes}). The main advantage of PET is the possibility to recognize and qualitatively measure the physiological disturbances before the morphological changes in tissues have appeared, while the X-ray or NMR computer tomography let one observe only the morphological changes. Thus, PET is a unique instrument for early diagnostics of cancer and other illnesses.

In the last decade a big progress in the field of PET technologies is achieved (see e.g.,~\cite{JinyiQi} and references there). Compact accelerators for USLI production are developed and produced, that provides the opportunity of wide dissemination of PET. At present there are about 300 educational, research and medical diagnostic PET centers. Unfortunately, there are still no PET installations in Ukraine. At the same time the cancer case rate is rather high in Ukraine especially among the population of the regions around Chernobyl and the people who took part in the accident elimination. 
Considering the high costs and complexity the first PET diagnostic complexes in all countries were created on the basis of functioning accelerators. This way one third of the whole cost was saved; besides for the maintenance of PET a group of nuclear engineers with necessary experience in this field was involved.
In such a way the PET diagnostic complex creation in Ukraine was approved on the meeting of the Ministry of Science in 2001.

National Science Center “Kharkov Institute of Physics and Technology” (NSC KIPT) has several accelerators, which can be used for production of USLI for PET~\cite{Bochek:1999,Dovbnyа:VANT}. Furthermore, NSC KIPT has a big experience on developing and creating detecting systems based on silicon photodiodes, the strip-detectors for the detector ALICE (CERN)~\cite{Kulibaba}, multiplicity detectors creation~\cite{Bochek:2001} etc. Taking into account the aforesaid the project of relatively cheep PET (first of all for the mammography - PEM) creation in NSC KIPT is proposed~\cite{Dovbnyа:URJ}.

The first step for the PET creation is choosing the optimal geometry of a detecting system for the compact PET (PEM) and developing corresponding software for the simulation of all its systems and for their easy operation.

%
%%%
%%%%%%%%%%%%%%%%%%%%%%%%%%%%%%%%%%%%%%%%%%%

\section{\label{sec:parameters} General features and parameters of PET}

The principle of PET operation is based on the reproduction of a 3D image of the radiation source by registering on coincidence two annihilated gamma-quanta, which fly back to back from the annihilation point of positron. During the diagnostic procedure a patient receives a little amount of radioisotope farm-medication, which contains a little amount of $\beta^+$-active USLI with activity of about several millicuries. In a short period of time the farm-medication gets into the organ under consideration, where radiation and annihilation of a positron take place. 
The important characteristics of USLI are its life time, the maximal energy of the emitted positrons $\varepsilon_{\rm max}$ and the defined by it maximal $R_{\rm max}$ and average $<$$R$$>$ depth of a positron penetration in a matter with density $\rho$.

The maximal depth of positron penetration $R_{\rm max}$ that is measured in g/cm$^2$ is determined by the following approximate relations~\cite{Kuchling}:
\begin{eqnarray}
& R'_{\rm max}=&0.542
 \ \varepsilon_{\rm max}
     -0.133, 
   \ \varepsilon_{\rm max}=( 0.8\text{--}1)
    \\
& R'_{\rm max}=&0.11
 \left(
  \sqrt{1+22.4 \ \varepsilon_{\rm max}^2}
   -1
    \right),
     \ \varepsilon_{\rm max}=(1\text{--}3)
      \nonumber 
\label{eq:production reaction}
\end{eqnarray}
and is connected with the value $R_{\rm max}$ measured in cm by the relations $R'_{\rm max}=\rho \, R_{\rm max}$. Here $\varepsilon_{\rm max}$ is in MeV.
The value of an average positron track in the matter of the object under consideration $<$$R$$>$ determines the limit value of the spatial resolution of PET: the more the kinetic energy of a positron (the bigger is $<$$R$$>$), the worse is the spatial resolution of PET.

In Table~\ref{tab:isotopes} the values of half-life time $\tau$, radiated positron maximal energy $\varepsilon_{\rm max}$, as well as the value of maximal $R_{\rm max}$ and average $<$$R$$>$ depth of positron penetration in water for  some USLI used in PET~\cite{Wagenaar} are presented:

\begin{table}[tbh]
\caption{Characteristics of USLI used in PET.}
\begin{center}
\begin{tabular}{ c c c c c }
  \hline \hline 
  ~~~Isotope~~~&
  ~~~$\varepsilon_{\rm max}$~~~&
  ~~~$R_{\rm max}$~~~&
  ~~~$<$$R$$>$~~~&
  ~~~~~$\tau$~~~~~\\
  &MeV&mm&mm&min\\
  \hline	
  $^{11}$C & 0.960 & 3.9 & 1.1 & 20.4\\
  $^{13}$N & 1.198 & 5.1 & 1.5 & 10.0\\
  $^{15}$O & 1.732 & 8.0 & 2.5 & 2.04\\
  $^{18}$F & 0.633 & 2.4 & 0.6 & 110\\
  \hline
\end{tabular}
\end{center}
\label{tab:isotopes}
\end{table}

The influence of positron kinetic energy on spatial resolution of PET was experimentally investigated for nuclides $^{11}$C in~\cite{Kulibaba} by comparing it with 514~keV gamma-radiation of $^{85}$Sr. The obtained result $<$$R$$>\, \approx 1$~mm is in good agreement with the calculated data presented in Table~\ref{tab:isotopes}.

The most important characteristics of PET are the spatial resolution and the efficiency of registration of annihilated gamma-quanta. In the modern production-run PETs that are widely used in clinical investigations the spatial resolution is 5--6~mm and is limited more by the sizes of a separate element of the detecting system and by photon statistics, than by the physical limitations caused by photon-photon noncollinearity and a positron track length. That is why practically the only reserve for improvement of spatial resolution is to decrease the sizes of the detectors which are used in PET scanners, since the increase of photon statistics due to the increase of activity of the radioactive farm-medication, which is administered to the patient, is undesirable because of increase of the radiation dose.

In the existing PET detecting systems scintillator and photoelectron multiplier are commonly used (see e.g.,~\cite{JinyiQi}). The use of photoelectron multiplier has two essential disadvantages: it demands high voltage and does not allow decreasing the value of the detecting element, what is necessary for improvement of spatial resolution of PET. That is why intensive investigations of a possibility to create detecting systems of PET on the basis of multi-element detecting modules, for example, with the use of siliceous photodiodes are being carried out recently~\cite{Lecoq}.

In the early PETs NaJ crystals were used as a scintillators. Soon they were replaced by BGO (bismuth germanate) crystals which have a higher efficiency of registration and are not hygroscopic. The maximal spatial resolution ability of PET that uses BGO has achieved approximately 4 mm.

Recently BGOs were replaced by new LSO (lutetium oxyorthosilicate) crystals~\cite{Huesman}. LSO has the registration efficiency similar to BGO, but with five times exceeding of light yield and eight times less time of highlighting. This allows obtaining spatial resolution ability close to its physical limit 1--2~mm, which is determined by an average track of a positron in a tissue and a little noncollinearity of annihilated photons.

\begin{figure}[tbh]
\begin{center}
\includegraphics[width=0.27\textwidth]{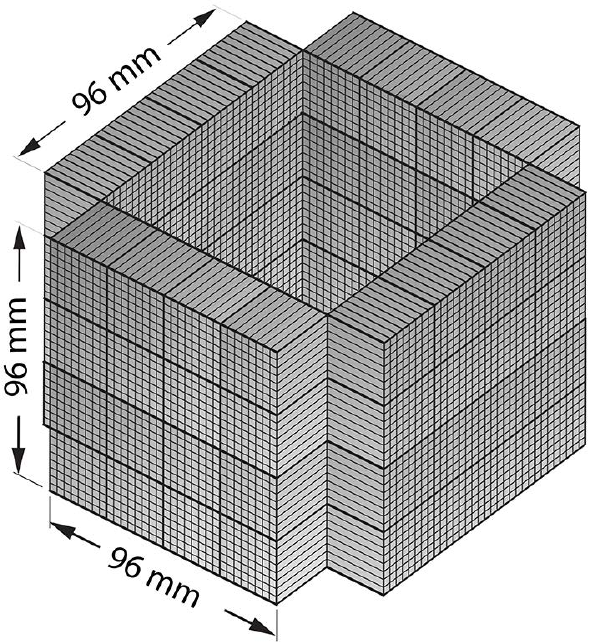}
\end{center}
\caption{Detecting system geometry.}
\label{fig:geometry}
\end{figure}

Thus, the detecting system determines the basic characteristics of PET~\cite{Globus}. The detecting system consists of the separate detecting elements, readout electronics, trigger system and data handling system. 
The scintillator and the siliceous photodiode constitute the detecting element of the prospective detecting system. The detecting system is constructed out of a big number of separate detecting elements that are placed around the working area of PET. It takes, for instance, the shape of a cube or a parallelepiped with two open facets (Fig.~\ref{fig:geometry}).

Multichannel electronics ensures reading-out analog signal from any detecting element and leading-out signals in one line. The trigger system provides a possibility to register two gamma-quanta on concurrence.

To choose the optimal geometry of the multi-element detection system of the compact PET based on the scintillator-photodiode type of detecting elements the original computer code is created. The program consists of three main blocks: simulation of the physical processes in the analyzed object, simulation of detection process and image reconstruction, which are described in the following sections.

%
%%%
%%%%%%%%%%%%%%%%%%%%%%%%%%%%%%%%%%%%%%%%%%%

\section{\label{sec:object} Simulation of analyzed object}

The analyzed 3D object is assigned by the set of inequalities:
\begin{equation}
\left\{ \begin{array}{c}
z \geq z_0-f_i(x,y)\\
z \leq z_0+f_i(x,y)
 \end{array} \right.
\end{equation}

We use the Monte-Carlo method for the simulation of diffusion processes of different neutron-lack radionuclides and $\beta^+$-decayed positrons in the analyzed object. From medical sources it is known that the background activity of normal tissues is about 5\,\% from the usual activity of the looked for objects.

Firstly we collect the points that have got into the volume of the assigned 3D object. Then the background is introduced with 5\,\% activity of the assigned object. We assume that this background is nearly uniform in the whole volume of the PET camera (see Fig.~{\ref{fig:compensation}a).

For the simulation of positron annihilation we draw the flying-out angle of two gamma-quanta as related to the laboratory system of coordinates (noncollinearity of gamma-quanta fly-out is not taken into account). Distribution on the flying-out angel was simulated isotropically over the solid angle.

%
%%%
%%%%%%%%%%%%%%%%%%%%%%%%%%%%%%%%%%%%%%%%%%%
\onecolumngrid

\begin{figure}[tbh]
\begin{center}
\includegraphics[width=1\textwidth]{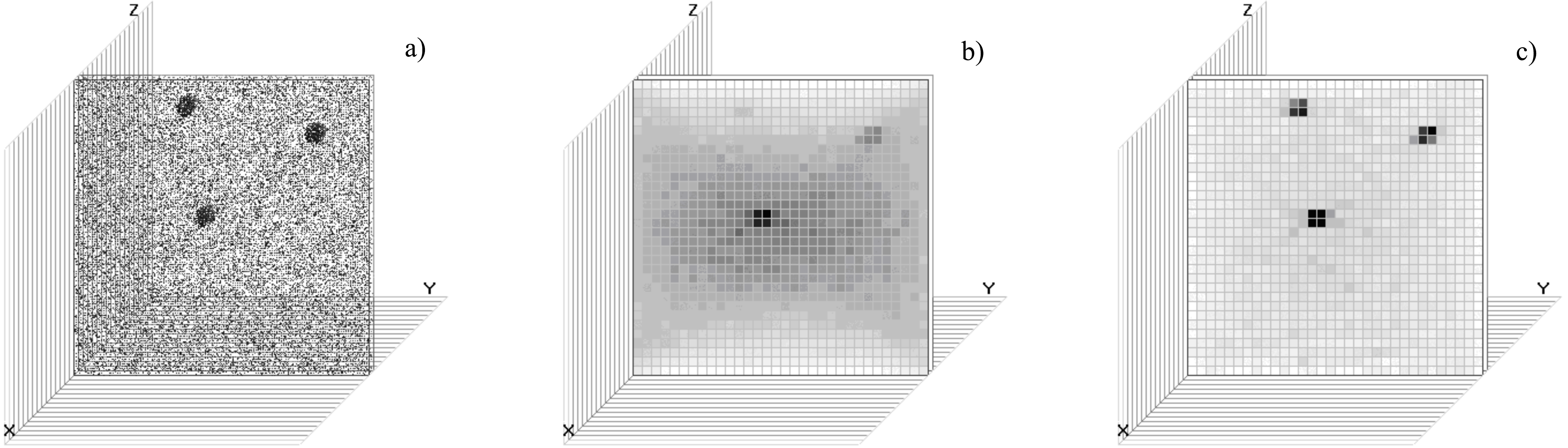}
\end{center}
\caption{The illustration of the opened boarders effect (upper and bottom boarders of the PET camera are opened): \\
a)~simulation of assigned objects (three balls of  6~mm diameter each) and the background; b)~the reconstructed image without the opened boarders compensation procedure; c)~the  reconstructed image with the compensation procedure.}
\label{fig:compensation}
\end{figure}

%\newpage

\twocolumngrid
%%%%%%%%%%%%%%%%%%%%%%%%%%%%%%%%%%%%%%%%%%%
%%%
%

%
%%%
%%%%%%%%%%%%%%%%%%%%%%%%%%%%%%%%%%%%%%%%%%%

\section{\label{sec:detection} Simulation of detection process}

The second block of the program simulates the process of detection of annihilating gamma-quanta. We assume that the PEM detection system has a rectangular geometry with four internal surfaces covered by 3$\,\times\,$3~mm$^2$ detecting elements and two opposite opened facets. 

Detection is carried out the following way. We get the annihilation point $(x_a,y_a,z_a)$ and the flying-out angles from the previous block.
Using these data we write an equation of the line in a three-dimensional space:
\begin{equation}
\left\{ \begin{array}{r c l}
(x_a-x)&=& r \,  {\rm cos} \varphi \,  {\rm sin} \theta, \\
(y_a-y)&=& r \,  {\rm sin} \varphi \,  {\rm sin} \theta, \\
(z_a-z)&=& r \,  {\rm cos} \theta
 \end{array} \right.
\label{line}
\end{equation}

This is a system of three equations with four variables: $(x,y,z)$ is a point on the line and r is the distance between $(x,y,z)$ and $(x_a,y_a,z_a)$.
By adding in turn the equations of the detecting planes  $(x=0, y=0, x=x_{\rm max}, y=y_{\rm max})$ to the given system of equations and by checking if other coordinates get into the region of detectors we find the points of crossing of the given line with the detecting planes. 

Thus, we determine coordinates of the two detecting elements that have registered the event (worked on coincidence). It is assumed that the detecting elements register events with 10\% efficiency.

%\smallskip

%
%%%
%%%%%%%%%%%%%%%%%%%%%%%%%%%%%%%%%%%%%%%%%%%

\section{\label{sec:reconstruction} Image reconstruction}

For image reconstruction we use the ``direct contrast'' method. It is based on filling the 3D matrix of elements of internal space of the PET camera the number (brightness level) of crossings of this element by straight lines connecting the pairs of detectors in the gate circuit. To get the final image we use exponential filtering:
\begin{equation}
F(x,a)=
 \frac{  \ e^{(1-x)\,a}-e^{-a}\ }
        {1-e^{-a}},
\label{eq:contrast}
\end{equation}
where $x$ is the normalized input level of brightness, $a$ is the contrast factor.

The opened boarders of the detecting system lead to decreasing efficiency of registration of events in the region close to the opened border. To compensate this effect we use the special matrix of estimated efficiency of detecting the annihilation event for each element of internal space of the PET camera. This efficiency is determined by the relation of the solid angle, at which the element is observed by all the detection elements under the coincidence condition, to the whole solid angle.

One can see the result of using this method when comparing the reconstructed images before and after the compensation procedure (see Figs.~\ref{fig:compensation}b,c).

The result of image reconstruction is represented in Fig.~\ref{fig:interface}. The analyzed object is composed of eight balls placed in tops of the cube with 2~cm side. The reconstructed image (right-hand) shows the drawback of this method, which reveals itself in the interference problem of close objects. The next step of our investigation will be to implement a more effective image reconstruction algorithm into the computer code (for example,~\cite{Herman,Gruzman}).

%
%%%
%%%%%%%%%%%%%%%%%%%%%%%%%%%%%%%%%%%%%%%%%%%

\section{\label{sec:interface} Interface description}

The interface of the PET simulator program is presented in Fig.~\ref{fig:interface}.

To carry out tests one can set 3D objects of various structure (Object combo) and use various isotopes (Isotope combo): $^{11}$C, $^{13}$N, $^{15}$O, $^{18}$F. Statistics can also be changed (Statistics combo).

Review of the reconstructed images of 3D objects is carried out in layers along the $Ox$ axis (Layer scroll). To estimate the accuracy of reconstruction procedure a possibility of overlap of the origin image of the object and the reconstructed image is provided (Show Object check).

%
%%%
%%%%%%%%%%%%%%%%%%%%%%%%%%%%%%%%%%%%%%%%%%%

\section{\label{sec:conclusions} Conclusions}

As a result of this work a computer program for modeling the multi-element detecting system of positron-emission tomograph was created, which allows: optimizing the construction of separate detecting elements and the prototype of the detecting system for concrete tasks of positron-emission tomography; conducting the efficiency tests of various algorithms of image reconstruction; studying the influence of various factors, such as geometric peculiarities of the detecting system, efficiency of gamma-quanta registration, background distribution of radioactive isotopes in the considered object, positron diffusion, etc. on the positron-emission tomograph detecting system efficiency. 

The investigation of the influence of the open facets of the detecting system has been carried out and the compensation algorithm for this effect has been introduced.

%
%%%
%%%%%%%%%%%%%%%%%%%%%%%%%%%%%%%%%%%%%%%%%%%
\onecolumngrid

\begin{figure}[tbh]
\begin{center}
\includegraphics[width=0.73\textwidth]{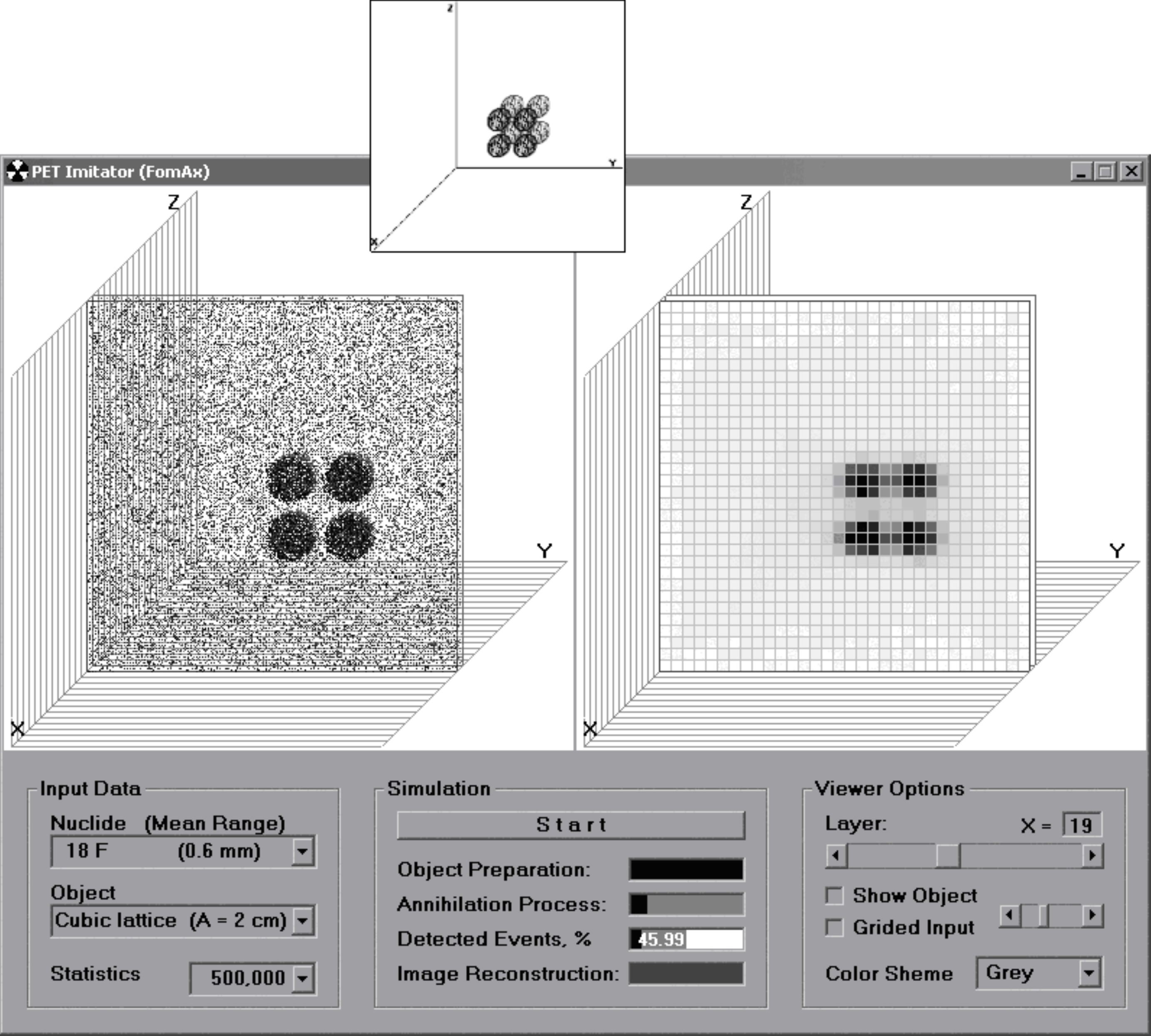}
\end{center}
\caption{Interface of PET simulation program.}
\label{fig:interface}
\end{figure}
\newpage
\twocolumngrid
%%%%%%%%%%%%%%%%%%%%%%%%%%%%%%%%%%%%%%%%%%%
%%%
%

%
%%%
%\newpage
%%%%%%%%%%%%%%%%%%%%%%%%%%%%%%%%%%%%%%%%%%%


\begin{thebibliography}{99}

\bibitem{Lecoq}
P.~Lecoq, P.~Le~Du., CERN Courier, November 2003.
% P. 29.

%\bibitem{Lawrence}
%Lawrence Berkeley National Laboratory site: lbl.gov

\bibitem{JinyiQi}
Jinyi~Qi {\it et al.},
IEEE Nuclear Science Symposium Conference Record {bf 4} (2001).
pubarchive.lbl.gov/islandora/
object/ir:118595/datastream/PDF/view

\bibitem{Bochek:1999}
G.L.~Bochek {\it et al.},
%Production of short-lived radionuclides on the linac “EPOS” NSC KIPT for PET.
Problems of Atomic Science and Technology. Ser.:NPI (Nuclear Physics Investigation), Kharkiv, NSC ”KIPT”
{\bf 1}, 33 (1999), (in Russian).
%p. 66-67

\bibitem{Dovbnyа:VANT}
A.N.~Dovbnyа, A.S.~Zadvorny, B.I.~Shramenko, 
%Production of short-lived radionuclides on the electron linac for PET.
Problems of Atomic Science and Technology. Ser.:NPI (Nuclear Physics Investigation), Kharkiv, NSC ”KIPT”
{\bf 3}, 34 (1999).
%Production of short-lived radionuclides on the electron linac for PET.
% p. 105-106.

\bibitem{Kulibaba}
V.I.~Kulibaba {\it et al.}
%Development and application of a silicon coordinate detectors
Problems of Atomic Science and Technology. Ser.:NPI (Nuclear Physics Investigation), Kharkov, NSC ”KhIPT”
{\bf 2}, 41 (2003), (in Russian).
% p. 85-88

\bibitem{Bochek:2001}
G.L.~Bochek {\it et al.},
%Silicon pad detectors for a simple tracking system and multiplicity detectors creation.
Problems of Atomic Science and Technology. Ser.:NPI (Nuclear Physics Investigation), Kharkiv, NSC ”KIPT” 
{\bf 1}, 37 (2001).
% p. 36-39.

\bibitem{Dovbnyа:URJ}
A.N.~Dovbnyа {\it et al.},
%The project of Kharkov regional PET Center,
Ukrainian Radiological Journal {3}, 34 (1999), (in Ukrainian).
% p. 316-318

\bibitem{Kuchling}
H.~Kuchling, Physik. Leipzig: “VEB Fachbuchverlag” (1980).
% 520 p.

\bibitem{Wagenaar}
D.J.~Wagenaar, 
med.harvard.edu/JPNM/physics/PETnucl.html  
Harvard: JPNM Physics (1996).

\bibitem{Huesman}
R.~Huesman {\it et al.},
IEEE Transitions on Medical Imaging {\bf 19} (2000).
% 532 p.

\bibitem{Globus}
M.E.~Globus, B.N Grunev, Anorganic scintillators. New and traditional matters. Kharkov: “AKTA” (2000), (in Russian).
% 408 p.

\bibitem{Herman}
G.T.~Herman, Image Reconstruction from Projections. The Fundamentals of Computerized Tomography. New York: Academic Press (1980).

\bibitem{Gruzman}
I.C.~Gruzman,
%Mathematical problems of computer tomography.
Soros’ educational journal
{\bf 7} (2001), (in Russian).
%117 p.

\end{thebibliography}
\end{document}